\documentclass[preprint,5p]{elsarticle}

\usepackage{amssymb}
\usepackage{soul}
\usepackage{xcolor}
\usepackage{url}
\usepackage{hyperref}

\definecolor{myOrange}{RGB}{255, 165, 2}
\definecolor{myGreen}{RGB}{78, 167, 46}
\definecolor{myBlue}{RGB}{30, 119, 180}
\definecolor{myPurple}{RGB}{126, 47, 142}

\biboptions{sort&compress}

\journal{npj Computational Materials}

\begin{document}

\begin{frontmatter}



\title{Describe, Transform, Machine Learning: Feature Engineering for Grain Boundaries and Other Variable-Sized Atom Clusters}



\cortext[cor1]{Corresponding author}

\author[BYU_CS]{C. Braxton Owens}
\author[LANL1]{Nithin Mathew}
\author[BYU_Phys]{Tyce W. Olaveson}
\author[NASA]{Jacob P. Tavenner}
\author[LANL1]{Edward M. Kober}
\author[Baylor]{Garritt J. Tucker}
\author[BYU_Phys]{Gus L. W. Hart\corref{cor1}}
\ead{gus.hart@gmail.com}
\author[BYU_ME]{Eric R. Homer\corref{cor1}}
\ead{eric.homer@byu.edu}

\affiliation[BYU_CS]{organization={Department of Computer Science Brigham Young University},
            city={Provo},
            postcode={84602}, 
            state={UT},
            country={USA}
            }

\affiliation[LANL1]{organization={Group T-1, Theoretical Division, Los Alamos National Laboratory},
            city={Los Alamos},
            postcode={87544}, 
            state={NM},
            country={USA}
            }
            
\affiliation[LANL2]{organization={Center for Non-Linear Studies, Theoretical Division, Los Alamos National Laboratory},
            city={Los Alamos},
            postcode={87544}, 
            state={NM},
            country={USA}
            }
            
\affiliation[NASA]{organization={KBR, Inc., Intelligent Systems Division, NASA Ames Research Center},
            city={Moffett Field},
            postcode={94035}, 
            state={CA},
            country={USA}
            }
            
\affiliation[Baylor]{organization={Department of Physics, Baylor University},
            city={Waco},
            postcode={76798}, 
            state={TX},
            country={USA}
            }

\affiliation[BYU_Phys]{organization={Department of Physics and Astronomy Brigham Young University},
            city={Provo},
            postcode={84602}, 
            state={UT},
            country={USA}
            }

\affiliation[BYU_ME]{organization={Department of Mechanical Engineering Brigham Young University},
            city={Provo},
            postcode={84602}, 
            state={UT},
            country={USA}
            }

\begin{abstract}
Obtaining microscopic structure-property relationships for grain boundaries are challenging because of the complex atomic structures that underlie their behavior. This has led to recent efforts to obtain these relationships with machine learning, but representing a grain boundary structure in a manner suitable for machine learning is not a trivial task. There are three key steps common to property prediction in grain boundaries and other variable-sized atom clustered structures. These are: (1) describe the atomic structure as a feature matrix, (2) transform the variable-sized feature matrices of different structures to a fixed length common to all structures, and (3) apply machine learning to predict properties from the transformed feature matrices. We examine these feature engineering steps to understand how they impact the accuracy of grain boundary energy predictions. A database of over 7000 grain boundaries serves to evaluate the different feature engineering combinations. We also examine how these combination of engineered features provide interpretability, or the ability to extract insightful physics from the obtained structure-property relationships.
\end{abstract}


\end{frontmatter}


\section*{Introduction}
\label{sec:sample1}
Due to the impact of grain boundaries (GBs) on material properties \cite{palumboApplicationsGrainBoundary1998, randleGrainBoundaryEngineering2010, watanabeGrainBoundaryEngineering2011a, rohrerGrainBoundaryEnergy2011b}, there is a need to better understand the relationship between the structure of a GB and its corresponding properties. With expanding computing power, increasingly large amounts of data, and advances in data-driven approaches, there has been a push for suitable representations of GBs in order to predict their properties \cite{echeverrirestrepoUsingArtificialNeural2014a, dangStandardDeviationEffect2023, nohDislocationDescriptorsLow2024, lortaraprasertRobustCombinedModeling2022, fujiiQuantitativePredictionGrain2020, priedemanQuantifyingConnectingAtomic2018a, kiyoharaPredictionInterfaceStructures2016a, zhangPredictingGrainBoundary2022a, zhuPredictingPhaseBehavior2018, yokoiNeuralnetworkInteratomicPotential2020, guziewskiMicroscopicMacroscopicCharacterization2021a, parsaeifardMaximumVolumeSimplex2020, cuiMachineLearningPredictionAtomistic2022c, homerMachineLearningInformedRepresentations2019a, trujillo2019machine, sharpMachineLearningDetermination2018, chesserLearningGrainBoundary2020a, wagihLearningGrainBoundary2020,wagihLearningGrainBoundarySegregation2022,hanGrainboundaryMetastabilityIts2016a,zhengGrainBoundaryProperties2020a, ratanaphanGrainBoundaryEnergies2015a, huGeneticAlgorithmguidedDeep2020a, tamuraFastScalablePrediction2017, gombergExtractingKnowledgeMolecular2017a, rosenbrockDiscoveringBuildingBlocks2017a, montesdeocazapiainCharacterizingTensileStrength2020, songAtomicEnergyGrain2022a, wuApplicationMachineLearning2020, nishiyamaApplicationMachineLearning2020a, yeUniversalMachineLearning2022a, snowSimpleApproachAtomic2019, huberMachineLearningApproach2018b}. However, accurate property prediction is not the only measure of success. Models and representations that provide insight into structure-property relationships are key to advance our understanding.
 
Representing a GB starts with defining its structure, since it has both macroscopic and microscopic characteristics. Macroscopically, five degrees of freedom are used to define the GB character: Three to define a misorientation between two crystals, (often given by a normalized rotation axis $[uvw]$ and angle $\theta$), and two to define a boundary plane $(hkl)$. Microscopically, the positions of the atoms result in $3n$ degrees of freedom. Additionally, a GB can assume various metastable configurations under any given set of macroscopic constraints.\cite{homerExaminationComputedAluminum2022a, zhuPredictingPhaseBehavior2018, hartGrainBoundaryPhase1972, frolovStructuralPhaseTransformations2013, hanGrainboundaryMetastabilityIts2016a, hickmanExtraVariableGrain2017a, meinersObservationsGrainboundaryPhase2020}. 

While there have been impressive developments in macroscopic representations for better understanding GBs \cite{zhangPredictingGrainBoundary2022a,yeUniversalMachineLearning2022a,ormeInsightsTwinningMg2016,francisGeodesicOctonionMetric2019a,zhouFirstprinciplesPredictionElectron2022,bairdFiveDegreeoffreedomProperty2021b}, the atomic structure is what defines a GB's properties. This article concentrates on the microscopic structure-property relationships since the macroscopic structure acts as a constraint on the microscopic structure.   

One microscopic method for defining a GB is the structural unit model \cite{suttonStructureTiltGrain1983,suttonStructureTiltGrain1983a,suttonStructureTiltGrain1983b,bishopCoincidenceLedgeDislocation1968}. This model describes the atomic structure of a quasi two-dimensional GB as a series of repeating atomic ``structural units" that are characteristic of the boundary's local atomic arrangement. This simplifies analysis of its geometry and properties as long as it is quasi two-dimensional. In recent years, the structural unit model has been modified to more accurately represent a GB by considering the effects of metastable structures \cite{hanGrainboundaryStructuralUnit2017}. Other early methods for defining local atomic environments for characterizing GBs include: the centrosymmetry parameter (CSP) \cite{kelchnerDislocationNucleationDefect1998}, Voronoi index \cite{zydekDescriptionGrainBoundary2021}, excess volume \cite{elliott2012introductory}, common neighbor analysis (CNA) \cite{honeycuttMolecularDynamicsStudy1987}, the Polyhedral Unit Model \cite{banadakiThreedimensionalPolyhedralUnit2017a}, and local entropy \cite{lejcekEntropyMattersGrain2021}. 

\begin{figure*}[t]
    \centering
    \includegraphics[width=1.4\columnwidth]{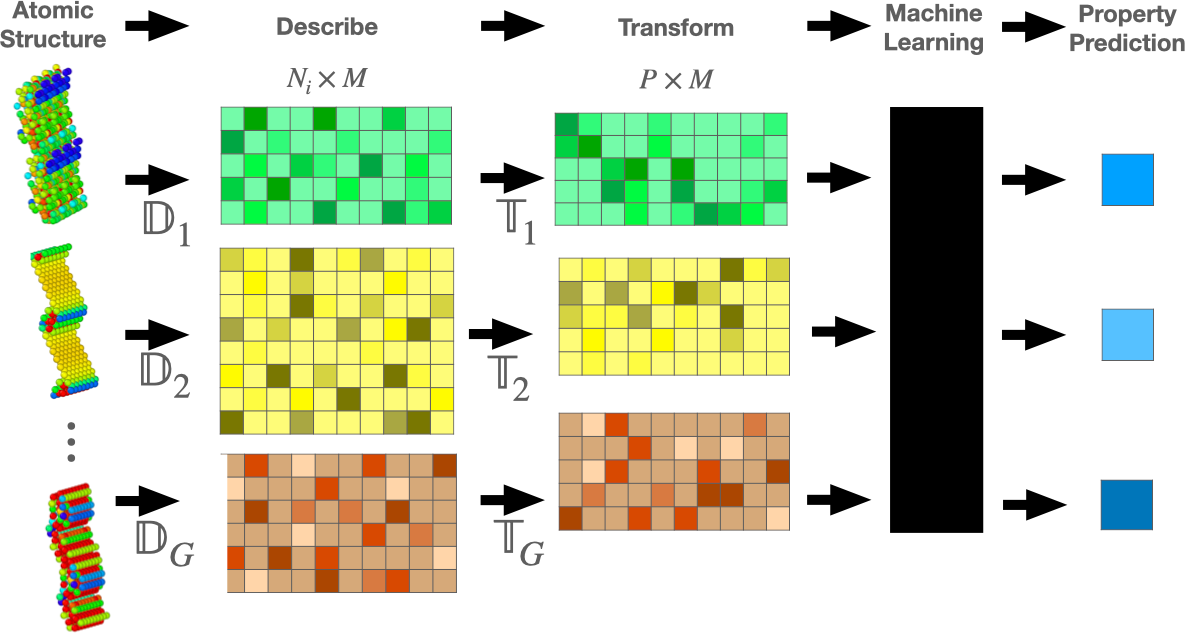}
    \caption{The workflow for predicting material properties from variable-sized atomic structures. Starting with various atomic structures (left), each structure is described using a matrix of size $N_i \times M$ that captures relevant features. These matrices are then transformed into a different representation of size $P \times M$. The transformed matrices are input into a machine learning model, which predicts the material properties (right), shown as blue squares. This approach allows for systematic analysis and prediction of properties based on the atomic-level description of materials.}
    \label{fig:flow}
\end{figure*}

Although the early methods have been used mainly to classify GB atoms \cite{gombergExtractingKnowledgeMolecular2017a, snowSimpleApproachAtomic2019, tsuzukiStructuralCharacterizationDeformed2007} they have also been used for machine learning predictions of atomic level properties \cite{huberMachineLearningApproach2018b,songAtomicEnergyGrain2022a}. Advanced methods, some of which are described below, have also been used to predict atomic level properties in GBs \cite{zhouChemomechanicalOriginHydrogen2016a}.

However, when using machine learning to predict properties of a GB as a whole, the descriptor must be transformed in some way to achieve a consistent feature size. This notion of transformation is an important one for variable-sized atomic structures such as GBs because different GBs will have different numbers of atoms (features) in their structure. Transformation is one step in a three step feature engineering process common to almost all machine learning predictions for variable-sized atom structures, such as GBs. 
These steps are illustrated in Figure \ref{fig:flow} and are described as follows:
1. \textbf{Describe} the atomic structure with an encoding algorithm, descriptor, or fingerprint of some kind, which is often represented as a matrix or vector. 2. \textbf{Transform} the mapping of the variable length descriptor for each structure to a fixed length descriptor common across all structures in a dataset. 3. Apply \textbf{Machine Learning} models or regression algorithms, to learn and then predict the property of a given atomic structure. 

We illustrate the consistency of these steps, occasionally combined in different orders or sequences, in numerous applications of machine learning to predict properties based on microscopic GB structure in the paragraphs below. These examples will also serve to highlight the variety of descriptors, transforms, and machine learning algorithms employed by the community.

Snow et al.\ \cite{snowSimpleApproachAtomic2019} utilized the graphs underlying CNA to \textbf{describe} the atoms in each GB, categorizing them into 2205 distinct environments. They then performed principal component analysis to reduce these environments into 84 principal components, which constitutes a second \textbf{description} step. To standardize the representation size, a \textbf{transform} was applied, representing each GB by the proportions of the 84 components present. This transformed representation was used as input to a linear regression \textbf{machine learning} model to predict GB energy.

Guziewiski et al.\ \cite{guziewskiMicroscopicMacroscopicCharacterization2021a} also explore this concept of proportions, utilizing the diamond-structure identification and the polyhedral template matching algorithms to \textbf{describe} GB atoms. This was \textbf{transformed} into a fixed-length density metric for each GB by counting the number of atoms within each polyhedral template class and normalizing this by the GB area. Random forest \textbf{machine learning} models were then used to predict both the GB energy and the tensile strength of the GB. 

Gomberg et al.\ \cite{gombergExtractingKnowledgeMolecular2017a} utilized a specialized pair correlation function \cite{billingeCrystallographyStudyDisorder2004, kalininBigDeepSmart2015} to \textbf{describe} their GB structures. This function is unique due to its use of a probability distribution function, allowing equal sampling for each GB and simultaneously \textbf{transforming} the descriptor into a fixed length. This representation was further refined to the first two principal components, constituting a second \textbf{describe} step, which is then used as input into a regression \textbf{machine learning} model. Dang and Yu extended Gomberg's method by incorporating the standard deviation of the pair correlation function through a weight parameter \cite{dangStandardDeviationEffect2023}.

More recently, the GB community has used atomic structure descriptors developed by the machine learned interatomic potential community. These descriptors are attractive because they are inspired by the symmetries and physical response inherent to the atoms, as described by Musil et al.\cite{musilPhysicsInspiredStructuralRepresentations2021a}.
Rosenbrock et al.\ \cite{rosenbrockDiscoveringBuildingBlocks2017a} implemented one of these physics inspired descriptors, called the smooth overlap of atomic positions (SOAP), to \textbf{describe} the GB and then \textbf{transformed} the SOAP descriptor into a fixed-length vector by averaging over the atom environments. These are then used to predict GB energy using a support vector \textbf{machine learning} model. Later, Fujii et al.\
\cite{fujiiQuantitativePredictionGrain2020} used SOAP to calculate a local distortion factor, \textbf{describing} how similar a GB atom's environment is to a bulk atom's environment. This was then \textbf{transformed} into a fixed-length using complete-linkage clustering. The GB thermal conductivity was then predicted by a ridge regression \textbf{machine learning} algorithm.

\begin{figure}[t]
    \centering
    \includegraphics[width=\columnwidth]{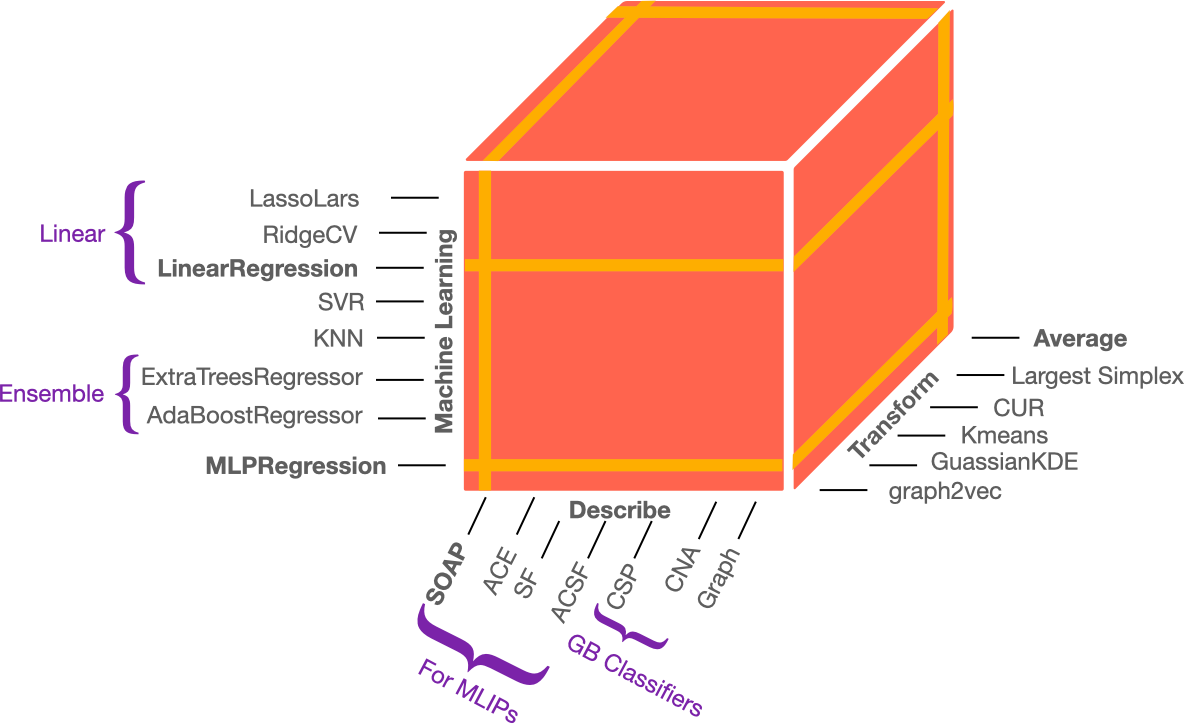}
    \caption{Graphic illustrating the various combinations of descriptors, transforms, and machine learning algorithms employed. The orange planes denote cross-sections of this space that are examined in this work. }
    \label{fig:cube}
\end{figure}

In this paper, we examine the impact of using different descriptors, transforms, and machine learning models for feature engineering for predicting GB energy, as illustrated in Figure \ref{fig:flow}. The exact \textbf{descriptors}, \textbf{transforms}, and \textbf{machine learning} algorithms we employ are listed in Figure \ref{fig:cube} and described in detail in the Methods Section. The feature engineering is tested on a dataset comprising 7304 aluminum GBs, which provides comprehensive coverage of the 5-dimensional macroscopic space of crystallographic character \cite{homerExaminationComputedAluminum2022a,Homer:2022:AlGBdataset}. The interplay of various descriptors, transforms, and machine learning algorithms are analyzed for their effect on the accuracy of their predictions. In addition, we examine how feature reduction impacts the results for select cases. Finally, we assess the interpretative ability of some key descriptors to establish meaningful connections to the inherent structure of the GB. Prioritizing interpretability is imperative to increase our understanding of atomic GB structure-property relationships.

\section*{Results}

\subsection*{Accuracy of Predictions}

As illustrated in Figure \ref{fig:flow}, there are essentially 3 methods or ``knobs'' that can be adjusted to improve accuracy of predictions: \textbf{descriptor}, \textbf{transform}, and \textbf{machine learning} algorithm. As noted above, the descriptors, transforms, and machine learning algorithms we employ are listed in Figure \ref{fig:cube} and described in detail in the Methods Section. 

\begin{figure*}[t]
    \centering
    \includegraphics[width=\textwidth]{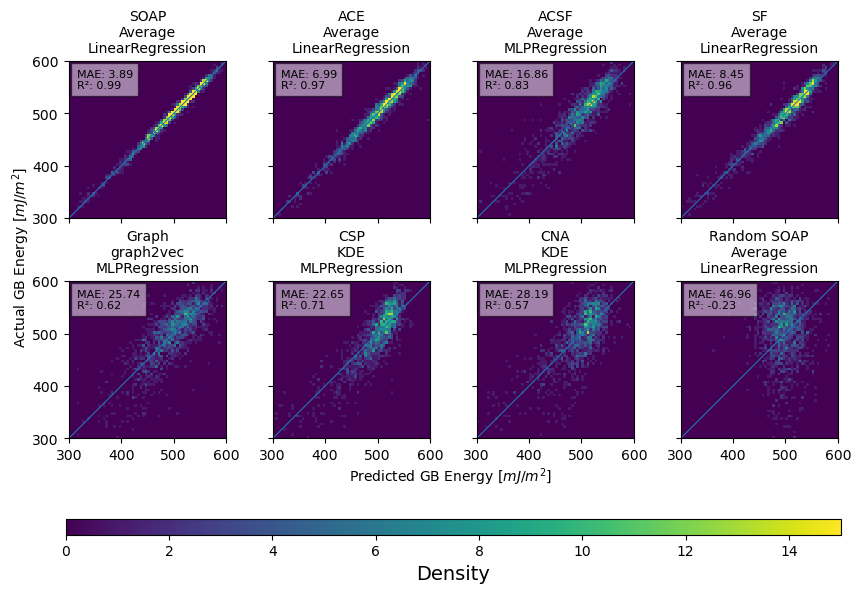}
    \caption{Each subplot showcases a different GB descriptor paired with its optimal transform and machine learning algorithm. The plots illustrate the accuracy of predictions across descriptors, using color-coded density levels to highlight the relationships between predicted and actual values. MAE is in units of mJ/m\textsuperscript{2}.}
    \label{fig:parity}
\end{figure*}

We start by examining Figure \ref{fig:parity}, which shows parity plots comparing machine learning predictions of GB energy against the GB energy values calculated in LAMMPS, as reported in \cite{homerExaminationComputedAluminum2022a}. There is one plot for each of the 7 descriptors examined in this work: The Atomic Cluster Expansion (ACE), SOAP, Atom Centered Symmetry Functions (ACSF), Strain Functional (SF) descriptors, a Graph description, CSP, and CNA. For each of these 7 descriptors, the results are reported for the combination of transform and machine learning algorithm that resulted in the highest overall accuracy predicted for GB energy using this dataset. The accuracy of predictions is measured by the mean absolute error (MAE) and R-squared ($R^2$) values. These metrics are used in tandem to best illustrate the performance of a given model. Finally, these 7 descriptors are accompanied by a parity plot labeled as ``Random SOAP'' where the SOAP descriptor is used as input, but with the GB energy values shuffled so that every SOAP vector points to a random GB energy value in the set. This serves as a worst-case reference value since a shuffled dataset would be expected to have no correlations.

In examining Figure \ref{fig:parity}, the SOAP descriptor combined with Linear Regression achieves the highest accuracy, with a low MAE of 3.89 mJ/m\textsuperscript{2} and a high $R^2$ of 0.99, indicating near-perfect correlation between predicted and actual values. In contrast, the Random SOAP model, where GB energies are shuffled, has a high MAE of 46.96 mJ/m\textsuperscript{2} and a negative $R^2$ of -0.23, confirming no predictive capability. While ACE and SF descriptors also achieve high accuracy, ACSF exhibit intermediate performance and descriptors like graph (graph2vec), CNA, and CSP exhibit significantly higher MAE and lower $R^2$, indicating poorer predictive performance. This suggests that the higher complexity descriptors do indeed capture more relevent information for predicting grain boundary energy.

Figure \ref{fig:parity} illustrates key aspects of feature engineering used to predict GB energy. First, it can be seen that due to the nature of this dataset, with many GBs concentrated about the mean GB energy value of 497 mJ/m\textsuperscript{2}, it is possible to obtain a relatively low MAE, even in the case of the ``Random SOAP'' model. It is for this reason that we report both the MAE and the $R^2$ values. Caution must be exercised in assuming a model is good just because the MAE is low. One must also see a high $R^2$ value to show that the model results in correlated predictions; a negative value for the $R^2$ metric indicates that it would have been better to simply predict the mean.

Second, one can see that the `average' transform is selected as the transform providing the most accurate predictions in four of seven cases. Third, in three of seven cases, the machine learning algorithm that provides the highest accuracy is linear regression. In the other four of seven cases, MLPRegression is the most accurate. But the three cases with linear regression have much better predictions than those with MLPRegression. Fourth, one can see that the combination of MAE and $R^2$ values provide a nice summary of the accuracy that can be visibly seen in the parity plots. Fifth, the stark contrasts of the MAE and $R^2$ values between the SOAP and ``Random SOAP'' models illustrates that there is valuable information in the features of the averaged SOAP that is correlated with the GB energy of a given structure. This is strictly true when comparing SOAP and ``Random SOAP'' and likely true when comparing ``Random SOAP'' with the other descriptors which categorize the atomic information of the GBs in distinct manners.

\begin{figure*}[t!]
    \centering
    \includegraphics[width=\textwidth]{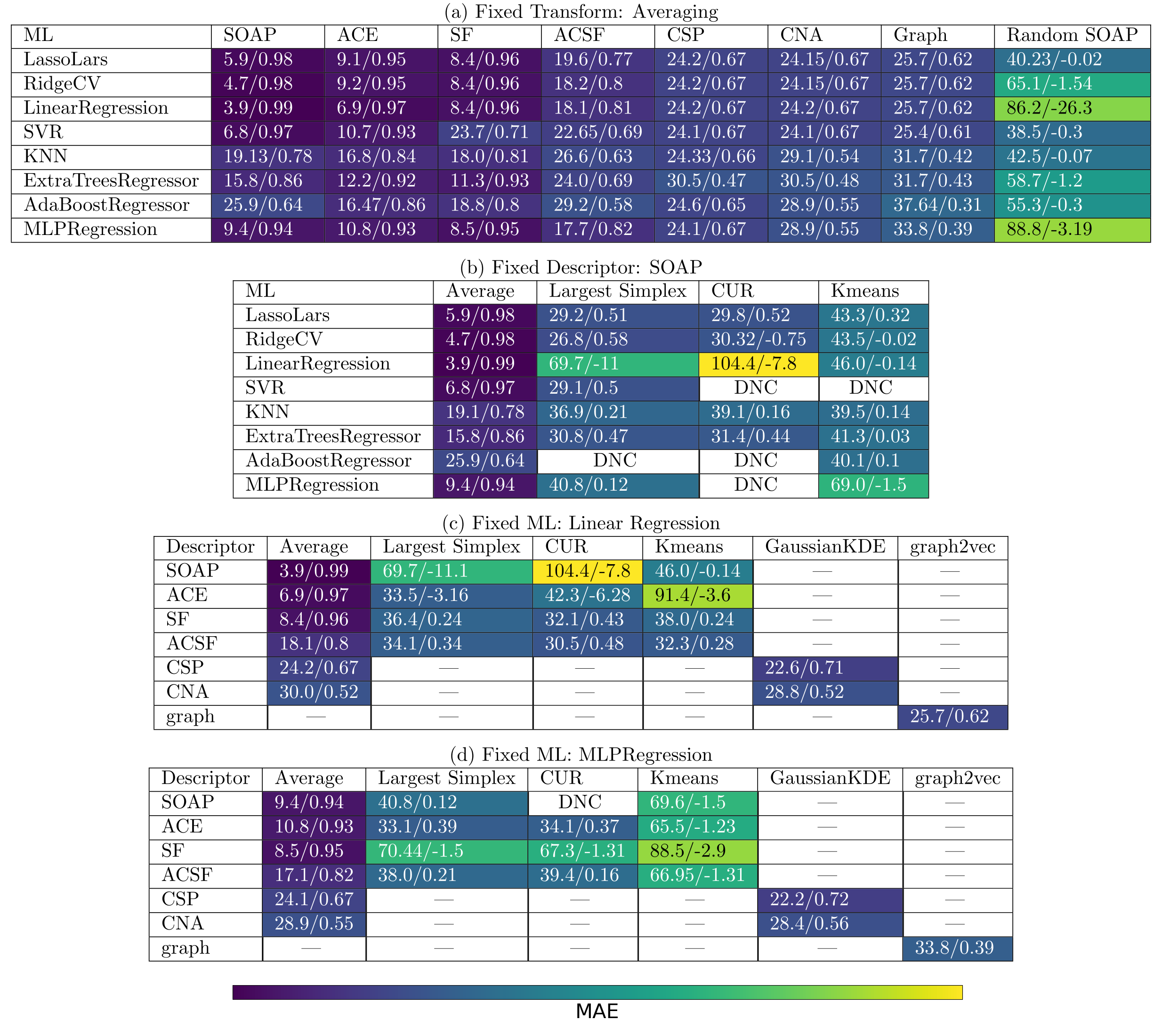}
    \caption{Table comparing descriptors and machine learning techniques, where the AMR transform is used in all cases except the \mytextsf{graph2vec} results. The entries in each cell list the MAE/$R^2$ as a measure of the accuracy (with MAE units in mJ/m\textsuperscript{2}), and the cell is colored according to this accuracy. Note that the \mytextsf{graph2vec} doesn't use the AMR transform because it has its own transform but is included here for comparison. The Random SOAP is included for reference to illustrate the accuracy on a randomized list of GB energies; values near this accuracy are considered to be no better than predicting a random output. If the models did not converge in a reasonable time their results were left as blank cells. Approaches that did not converge are labeled with DNC.}
    \label{fig:all_tables}
\end{figure*}

While important insights can be gained from these comparisons in Figure \ref{fig:parity}, great care must be exercised because they are not direct comparisons; the models are different. Better comparisons can be made by holding as many variables constant between the models as possible. The three key steps (\textbf{describe}, \textbf{transform}, and \textbf{machine learning}) represent a 3-dimensional space of combinations, as illustrated in Figure \ref{fig:cube}. However, since not all combinations were evaluated, we choose to examine 2-dimensional subsets, holding constant one of the three methods, as illustrated by the yellow bands in Figure \ref{fig:cube}. The methods that are held constant are chosen because they typically preform better than their counterparts for the GB energy predictions examined in this work. 

The first subset we analyze compares different descriptors and machine learning algorithms, keeping the averaging transform constant since it performes well for 4 of 7 descriptors in Figure \ref{fig:parity}. The accuracy comparison for this subset is given in Figure \ref{fig:all_tables}a. The ACE, SOAP, and SF descriptors performed exceptionally well across multiple machine learning algorithms. The $R^2$ values were consistently above 0.95 in the majority of cases, indicating strong predictive capabilities, while the MAE values were comparatively low, $<10$ mJ/m\textsuperscript{2} in the majority of cases, reflecting accuracy in the energy predictions. ACSF had higher errors (MAE just less than 20 mJ/m\textsuperscript{2} and $R^2$ around 0.8). CSP, CNA, and the graph descriptor had errors around 25 mJ/m\textsuperscript{2} and $R^2 < 0.67$.

One might be tempted to make quick judgements about the quality of the different descriptors, but caution must be exercised because the number of features for each descriptor vary drastically, as shown by Table \ref{tab:descriptor_input} in the Methods section. Each descriptor forms a unique basis to represent an environment. For the SOAP, ACE, and SF techniques, the user specifies the desired order of radial basis functions, the degree of spherical harmonics, or the polynomial order to be included in the representation. Similarly, for the ASCFs, the user selects sets of 2- and 3-body functions. The graph used in combination with \mytextsf{graph2vec} creates a nearest-neighbor connected graph where the user defines a cutoff distance for the nearest neighbors. In contrast, the CNA and CSP descriptors are both singular-valued quantities for each atom, though it is noted that CNA is built on top of local graphs that could be used instead of the integer classification, as in \cite{snowSimpleApproachAtomic2019}.

In the cases where the user picks the level of expansion, the user can determine what cutoff can be used to obtain a desired level of accuracy; this trades computation time and memory required for a larger basis that hopefully captures more of the physics. For example, when we increased the number of ACSF features from 8 to 37, the $R^2$ value increased from 0.65 to 0.81 but it required significantly more computation time.
One could also compare the fact that ACSFs and SFs have approximately the same number of terms, 37 and 36 respectively, but the ACSFs perform worse than SFs.  This suggests that more or different interaction pairs are probably required to detect the structural features that determine GB energy with ACSF. In fact, with a greater understanding of the ACSF implementation, it is possible that one could obtain higher accuracy even with the same number of terms. 
Although we see that this method of ``covering your bases'' by increasing the number of features does quite well in terms of accuracy, there is power in a pre-training choice of basis based on physical beliefs about the material, which ideas are discussed in the interpretability section of the results.

What is remarkable about the two singular-valued descriptors, CSP and CNA, is that, despite being less accurate than the SOAP, ACE, SF, and ACSF descriptors, they still achieve respectable accuracy (c.f. Figure \ref{fig:all_tables}a).  
In fact, it is simple enough that we show the equation for the LR model, 
$ \gamma = 280.26 \times x_{\mathrm{CSP}} + 173.14$, 
where $\gamma$ is the GB energy and $x_{\mathrm{CSP}}$ is the average of the scalar CSP values for a given boundary. Similarly for CNA, the single coefficient linear function is 
$\gamma = 505.88 \times x_{\mathrm{CNA}} -370.71 $, 
where $x_{\mathrm{CNA}}$ is the average of the integer CNA values that refer to different structure types.

Finally, in Figure \ref{fig:all_tables}a, we include the graph descriptor, despite its use of a different transform, \mytextsf{graph2vec}. This descriptor performs worse than the CSP and CNA descriptors, even though it encodes more information than the singular-valued descriptors. However, as with the other descriptors, several parameters could be optimized for better predictions, including the cutoff distance of the graph, the weighting of the graph, and the selection of subgraphs in the \mytextsf{graph2vec} transform.

In summary, the effect of descriptor on accuracy in Figure \ref{fig:all_tables}a shows that in general, more features is better. The density-based descriptors, many of which are created for machine learned interatomic potentials, appear to be better at capturing the complex and intricate nature of the local atomic environments. 

In examining the role of machine learning algorithms for a given descriptor, one can see that the linear models (LR, LASSO, RidgeCV) generally outperform other types of machine learning models for nearly all descriptors when using the average transform (c.f. Figure \ref{fig:all_tables}a). The SVM (SVR) and Nearest Neighbor (KNN) models frequently rank next in performance, followed closely by Neural Network (\mytextsf{MLPRegression}) and ensemble (\mytextsf{AdaBoostRegressor}) methods. All these models perform better than the "Random SOAP" input, demonstrating their ability to identify meaningful correlations between the features and GB energy. 

The second subset we examine in the 3-D space holds the SOAP descriptor constant and varies the transform and machine learning methods, as illustrated in Figure \ref{fig:cube}. The results for this subset are illustrated in Figure \ref{fig:all_tables}b, where it should be noted that the average column in Figure \ref{fig:all_tables}b is the same as the SOAP column in Figure \ref{fig:all_tables}a due to the intersection of the two 2-D cross-sections. This analysis shows that in all cases in Figure \ref{fig:all_tables}b, the averaging transform significantly outperforms the other methods; MAE is less than 10 mJ/m\textsuperscript{2} for the average transform in most cases and greater than 29 mJ/m\textsuperscript{2} for the other transforms in all cases. 

All the transforms assume some prior on the important features to transform. The average transform assumes that an average environment is the most important information to preserve. \mytextsf{KMeans} clustering assumes that the clustering in the dataset and the locations of those clusters is the most important information to preserve. Largest simplex assumes that it is important to represent the data with subsections of the data that are far apart and maximize the simplex volume. CUR assumes that specific subsets, actual rows and columns of the original data, are critical and obtains these through matrix decomposition.

Given that GB energy is calculated from the sum of the energy of all the atoms divided by the area of the boundary, it is not surprising that the best tranform is an average of the atomic environments. In other words, the assumption behind the average transform aligns closely with the calculation method for GB energy, making it a suitable representation. Conversely, the assumptions underlying other transforms do not as accurately capture the relationship between the atomic collection and GB energy. However, there may be other cases where a different transform better matches with the property of interest. For example, in some cases extreme values of a distribution control the behavior, such as in fracture, and a different transform may better capture that relationship. Therefore, we hypothesize that the best transform for accurate predictions is one that preserves the relationship between the way a collection of atoms relates to the property of interest.

However, it is also possible that the choices made in this work about the transform hyperparameters resulted in poor predictions we did not seek to optimize these hyperparameters. For example, the target rank used in the CUR transform is 20 and the largest simplex transform employed 10 dimensions; it is not clear that these values are sufficient or insufficient. Similarly, the \mytextsf{KMeans} transform employed 100 clusters and it is not clear that this is representative of the number of clusters in any of the descriptors. 

It is also worth noting that the SOAP descriptor appears to fall victim to the curse of high dimensionality when transformed with \mytextsf{KMeans} clustering. On average, the \mytextsf{KMeans} clustering transform performs worse when applied to SOAP. This is likely because the high dimensionality of SOAP descriptors leads to a phenomenon known as ``distance uniformity". In high-dimensional spaces, feature values tend to be equidistant from each other, making it difficult for clustering algorithms to distinguish between similar and dissimilar data points. 

The third and fourth 2-D subsets 
we examine from the 3-D space illustrated in Figure \ref{fig:cube}
hold a different machine learning algorithm constant. Specifically, the linear regression model was picked because of its high accuracy and \mytextsf{MLPRegression} was also picked because of the popularity of deep learning models. The results for these subsets are presented in Figures \ref{fig:all_tables}c and \ref{fig:all_tables}d where the effect of different descriptors and transforms can be seen. First, in comparing the two tables, the average transform is better with linear regression in all but the case of ACSF. The higher accuracy between linear regression and \mytextsf{MLPRegression} is evenly divided for the largest simplex transform. \mytextsf{MLPRegression} is better in two of three cases for the CUR transform and in one of four cases for the \mytextsf{KMeans} transform. However, in many of these cases, the accuracy values approach or exceed predictions by ``Random SOAP'', making it difficult to judge the value of the improvements. Furthermore, these all perform worse than the average transform.  

In the case of CSP and CNA, the \mytextsf{GaussianKDE} transform performs better than the average transform for both linear regression and \mytextsf{MLPRegression}, with the exception of CNA by linear regression. Also for these, \mytextsf{MLPRegression} performs better than linear regression for three of four cases considered. In these singular-valued descriptors, the more sophisticated \mytextsf{GaussianKDE} transform and \mytextsf{MLPRegression} allow it to obtain slightly better predictions. 

This examination of the effect of all three key steps (\textbf{describe}, \textbf{transform}, \textbf{machine learning}) shows that the descriptor plays an outsize role in the quality of the predictions. However, the transform of the features also plays an important role and some important information can be lost at this step if care is not exercised. Finally, the machine learning algorithm appears to play more of a secondary role; if the features are correlated with the property of interest, multiple algorithms can often extract the relationship (though some methods appear to perform better than others depending on the circumstance).

\begin{figure*}
    \centering
    \includegraphics[width=1.4\columnwidth]{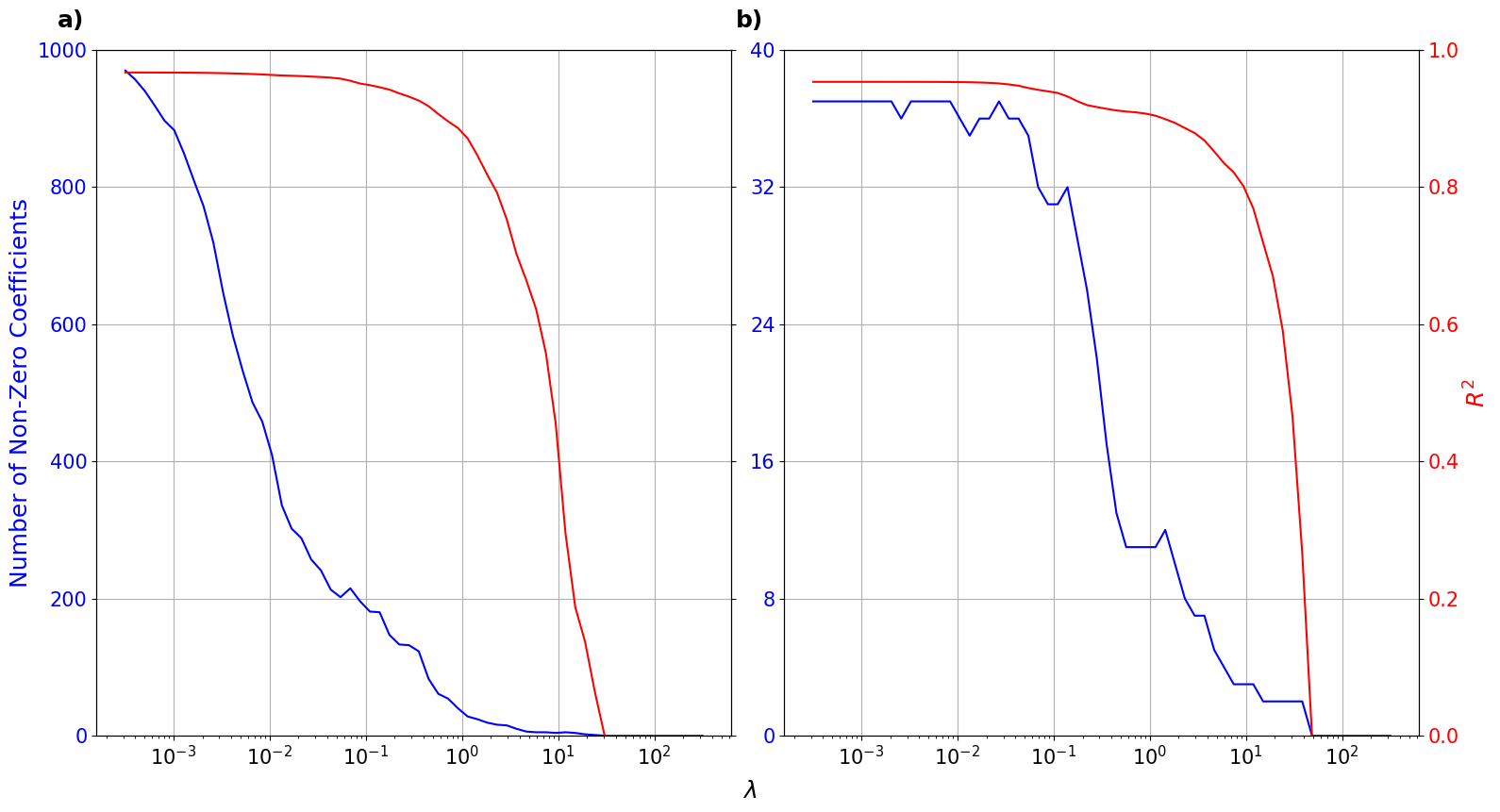}
    \caption{Figure of LASSO model trained on averaged SOAP (a) and SF (b). The x-axis is the $\lambda$ value that scales the regularization term. The regularization term is the L1 norm of vector of coefficients. Increasing $\lambda$ pushes the model to send coefficients to zero. The red line shows the $R^2$ value calculated from predictions of the model. The blue line represents the number of non-zero coefficients.}
    \label{fig:lasso_soap_sf}
\end{figure*}   

\subsection*{Feature Selection}

Although SOAP achieves the highest accuracy when predicting GB energy, in our implementation SOAP also uses the most features of any of the descriptors (c.f. Table \ref{tab:descriptor_input}). ACE and SF achieve comparable accuracy but only use 121 and 36 terms, respectively. In fact, these two occasionally outperform SOAP. It should be noted that any of these could be adjusted to use more or fewer terms to achieve higher or lower levels of accuracy. One can also use feature selection methods to remove redundant or irrelevant information in the machine learned structure-property models. Feature selection is an important step towards interpretable machine learning models because of the challenge of interpreting the meaning of high-dimensional descriptors \cite{Wright-Ma-2022}.

To identify lower dimensional representations we implement a feature selection method that uses the least absolute shrinkage and selection operator (LASSO) to identify what terms are most important for retaining high accuracy. LASSO is formulated as a minimization of a least squares error plus an L1 norm regularization term scaled by the parameter $\lambda$ which controls the trade-off between fitting the model accurately to the training data and keeping the model coefficients (parameters) small and sparse. This LASSO model is defined by

\begin{equation}
\min_{\beta} \left\{ \frac{1}{N} \|y - X\beta\|^2_2 + \lambda \|\beta\|_1 \right\}
\end{equation}

For this feature selection analysis, we use SOAP due to its prominence in GB predictions and SF for its accuracy and interpretability, as detailed in the following section. Figure \ref{fig:lasso_soap_sf} plots the number of non-zero terms(blue) and the model accuracy(red) of LASSO as the size of $\lambda$ increases for both SOAP (Figure \ref{fig:lasso_soap_sf}a) and SF (Figure \ref{fig:lasso_soap_sf}b). The sparsity of the parameters increases with increased $\lambda$ values. This model is trained using the average transform and illustrates just how many parameters can be neglected while maintaining a high $R^2$ value. The `elbow' of the $R^2$ curve marks where the model begins to significantly drop in accuracy.

Figure \ref{fig:lasso_soap_sf} illustrates that at low $\lambda$ values SF retains all 36 terms and achieves an $R^2$ value of $0.95$. SOAP starts with more terms but can be reduced to 209 terms with a comparable $R^2$ value. At the high end of $\lambda$ values, SOAP and SF achieve an $R^2$ value of $0.43$ using 4 and 2 terms, respectively,. 

This illustrates that although the accuracy decreases, both of these descriptors can be reduced to a very small representation space while preserving the most important information. It is noteworthy that averaged values of both CSP and CNA achieve higher accuracy predictions of GB energy with a single scalar value with $R^2$ values of $0.67$ for both. Thus, these two singular-valued descriptors are quite expressive and are better than just a few terms of the other descriptors. But, perhaps this is not surprising since these descriptors were designed to easily identify defects and other changes in structure with a single value, while the other descriptors were created to provide a more nuanced description of an environment with a much larger number of terms.  Consequently, while feature selection can be helpful in removing redundant and irrelevant information, the descriptors selected have a big impact on identifying important features that contribute to the machine-learned structure-property model.

\subsection*{Interpretability}

The goal in this work is to obtain structure-property models that are both accurate and interpretable. Interpretability is of critical interest for the advancement of science since machine learning models could easily become black box models that we can't understand. We have examined the impacts that the descriptor, transform, and machine learning model have on accuracy and methods to select the features that have the biggest impact on the models. We now focus on extracting interpretable information from these models. 

\paragraph{Descriptors}
ACE, SOAP, SF, ACSF, CSP, and CNA all represent the local environment around each atom. These representations often involve sums over neighboring atoms, where each term in the sum depends on the distance and/or angular relationship between the central atom and its neighbors. ACE, SOAP, and SF all represent the angular relationships in terms of spherical harmonic expansions, while ACSF uses a more explicit $n$ body expansion of the angular terms. SOAP utilizes a truncated Gaussian function for the radial distribution and then recasts its basis function in a bispectrum approach. This results in the mixing of angular and radial nodes in the invariant basis functions. ACE utilizes multiple radial functions within a specified cutoff radius to capture the radial environments. In SF, the radial and angular nodes of the expansion are kept separate resulting in descriptors that are analogous to an atomic orbital expansion. SF then represents this information in the minimal set of invariants required to characterise deformations up to the 4\textsuperscript{th} order (i.e., second derivatives or curvature of the strain). ACSF uses multiple Gaussian radial terms which are then convolved with the angular terms. All of the methods carry out their expansions to a level of accuracy that can be defined by the user.

In contrast to the other methods, the graph description (used in the \mytextsf{graph2vec} transform) characterizes the GB in a periodic, weighted graph. In this framework, the nodes of the graph correspond to the spatial positions of atoms and the edges are weighted by the distance between neighboring atoms within a cutoff distance. Thus, this method captures the complex arrangements of the atoms in a GB as a whole.

All the descriptors transform the 3D spatial coordinates of atoms into a high-dimensional feature space. This transformation is designed to capture complex interactions and symmetries, but it also means that the resulting descriptors are often far removed from the intuitive, three-dimensional space in which atoms actually exist. The use of physical descriptors (such as bond bending terms or spherical harmonics) leaves open the possibility that terms \textit{could} be interpretable. However, the derivation of the terms often renders this difficult, and most approaches do not offer any interpretations of their descriptors. This is not a criticism of these descriptors as they were not designed for the purpose of interpretability. 

Many of these descriptors were designed for machine learned potentials or for classification of atom neighborhoods. In other words, many of these descriptors were meant for forward modeling only, but to extract interpretable information, we need to go backwards through forward models. Therefore, pointing out weaknesses in backwards applications to extract specific terms is not a completely fair criticism. Nevertheless, because we wish to understand which descriptors provide this backwards path to interpretability we examine them anyway.

In a backwards application, ACE and SOAP are less than ideal because they do not retain knowledge of how their rotationally invariant descriptors are oriented in space. In contrast, the SF approach explicitly retains such terms. Additionally, for ACE and SOAP, one can know which degree of the spherical harmonics are important as illustrated in Supplemental Figure S2, 
but finer structural detail within that degree is lost. It may be possible to preserve some of that information if interpretability is desired. In a backwards application of ACSF, it is not clear to the authors how one would interpret the values of the 2- and 3-body functions. At a minimum the summation over neighbors makes it difficult to connect with specific atomic structures. However, it is possible that improved application of ACSF descriptors could be useful in extracting the local symmetries and deviations therefrom that could result in interpretability. 

\mytextsf{Graph2vec}'s interpretability suffers due to its abstract representation of graph embeddings. In its current implementation, there is no method by which to go backwards and make use of the subgraphs that were extracted, but this could be added for improved interepretability. CSP and CNA are non-unique descriptors where multiple atomic configurations can produce identical values, making it challenging to pinpoint specific structural features responsible for observed properties.

As a result, the inherent complexity of these descriptors makes it challenging to reverse-engineer the key features of the structure-property relationships. Most of these descriptors do not support easy interpretability because they were not designed for this purpose. However, the SF approach is distinct in characterizing the invariants of physical deformations up to the 4\textsuperscript{th} order. This descriptor is examined in additional detail in the case study below.

\paragraph{Transforms}
Just like descriptors, the ability to go backwards through transforms is crucial for interpretability. Averaging, the best transform for the most accurate descriptors, can not be applied backwards. One can only make conclusions about the average atomic environments. As explained in the Accuracy subsection, there are assumptions inherent to each of the transforms. Averaging best matches the physics of how GB energy is calculated. 

On the other hand, the largest simplex and CUR methods actually preserve specific features from the input matrix. By selecting the most important rows, these techniques effectively fix the size of the matrix that describes the GB. In other words, they choose the atomic environments that, even before training on data, are likely to preserve valuable information based on the assumptions of these transforms listed earlier. This deliberate selection ensures that the matrix representation of each GB remains concise and meaningful. Therefore, both CUR and largest simplex methods identify original atomic environments that can be analyzed for their interpretability and impact on the predictions.

The \mytextsf{KMeans} clustering transform, while fixing the matrix representation, does not preserve the original rows of the matrix but rather a number of cluster centers. Thus it is like averaging where information is lost in the transform application.  In the case of \mytextsf{KMeans} clustering the original rows can be identified by finding the nearest neighbors to the cluster centers. Although this can be difficult in higher dimensions. 

Finally, it is important to note however that other properties of interest might have different relationships with the atoms involved and therefore a different transform might work best, as discussed in the Accuracy subsection above. 

\paragraph{Machine Learning Models}
Finally, the ability to extract interpretable information from the machine learning models depends heavily on the model used. When linear models are used, one can readily identify the features that have the greatest impact on the predictions based on values of the coefficients. The relationships are easily defined and understood. One can even reduce the features using a feature selection or regularization approach \cite{Wright-Ma-2022}, as discussed above, to more easily identify the important features. However, in non-linear models identifying the most important features is not as simple. 

If a \mytextsf{KNN} performs well, that may be indicative of clustering in the input features, which would suggest that the clustering is relevant to the property of interest. Ensemble methods like \mytextsf{AdaBoostRegressor} fit the data multiple times, focusing more on difficult cases with additional iterations. While decision trees provide inherent interpretability, ensembles of them make that more difficult. However, these algorithms can export importance scores to learn about which features are of greatest interest.

In our Support Vector Regression (\mytextsf{SVR}) model, we utilized a linear kernel to fit the data with a hyperplane. The choice of a linear kernel allowed us to maintain a straightforward relationship between the features and the output. This approach ensured that the model remained easily interpretable, as each feature's effect on the output could be independently assessed through the corresponding coefficients. By employing a linear kernel, we avoided the complexities associated with high-dimensional transformations, which are common in nonlinear kernel methods.

A neural network model uses a series of layers whose connectedness and construction can be highly variable. In the neural network each layer typically has a non-linear activation function to learn non-linear relationships. While this may improve predictions, the overall structure of the neural network makes it difficult to extract interpretable information.

However, in any of these cases, one can use additional tools, such as those that fit into the category of Explainable AI \cite{longoExplainableArtificialIntelligence2024}, to extract the features of greatest impact or importance.

Having just discussed how the method selected for each step in the process of predicting structure-property relationships impacts interpretability, the discussion has remained theoretical. In the following subsection, we examine a case study where we can be more specific about the ability to extract meaningful information from machine learned models.

\subsection*{Case Study of Interpretability}

\begin{table}[t]
    \caption{Comparison of the top five features identified by LASSO and SHAP for the SF model. Each feature is ranked based on its influence on the model's predictions, where the $+$ or $-$ indicating a positive or negative influence, pushing the model's output higher or lower, respectively. This is accompanied by a column listing the features with the highest correlation with GB energy along with the sign of the correlation.
    }
    \label{tab:top5}
    \centering
    \begin{tabular}{cccc}
    \hline
Rank & LASSO & SHAP & Correlation\\ \hline
1 &  P4I8 $-$ &  P4I8 $-$ & P1I0 $+$ \\
2 &  P1I0 $+$ &  P2I0 $+$ & P3I4 $+$ \\
3 &  P2I0 $+$ &  P1I0 $+$ & P4I8 $-$ \\
4 &  P4I9 $-$ &  P3I4 $+$ & P2I2 $-$ \\
5 &  O2I0 $+$ &  P4I9 $-$ & P3I0 $+$ \\ \hline
    \end{tabular}
\end{table}


Here we examine how a combination of one descriptor, one transform, and features selection in two different machine learning models provide interpretability. We employ the SF descriptor because, as noted above, it was defined with the express purpose of retaining a physical meaning. We employ the average transform because, as noted above, it retains a connection to how our property of interest, GB energy, is calculated. Finally, we examine feature selection in two different machine learning methods to illustrate the differences related to interpretability.

As discussed above, regularized linear models allow easy identification of the most important features in a model. We revisit the results of the LASSO application to the SF illustrated in Figure \ref{fig:lasso_soap_sf} and described in the Feature Selection subsection above. As the model complexity is reduced, the terms that remain can be considered the most important for GB energy prediction and interpretability. The last five SF terms to be removed by LASSO, and therefore the top five terms for predicting GB energy, are listed in Table \ref{tab:top5}. Next to each term in Table \ref{tab:top5} is the sign of the correlation of that term in the model with its effect on the GB energy prediction.

To understand the impact of a non-linear model, we employ Extra Trees regression. For interpretability, this is used in conjunction with SHapley Additive exPlanations (SHAP) analysis \cite{NIPS2017_7062}. SHAP values explain the prediction of an instance by computing the contribution of each feature to the prediction. This method is based on game theory and helps in attributing the prediction output to individual features, offering a fair and consistent way to understand the model behavior.

\begin{figure}[t]
    \centering
    \includegraphics[width=\columnwidth]{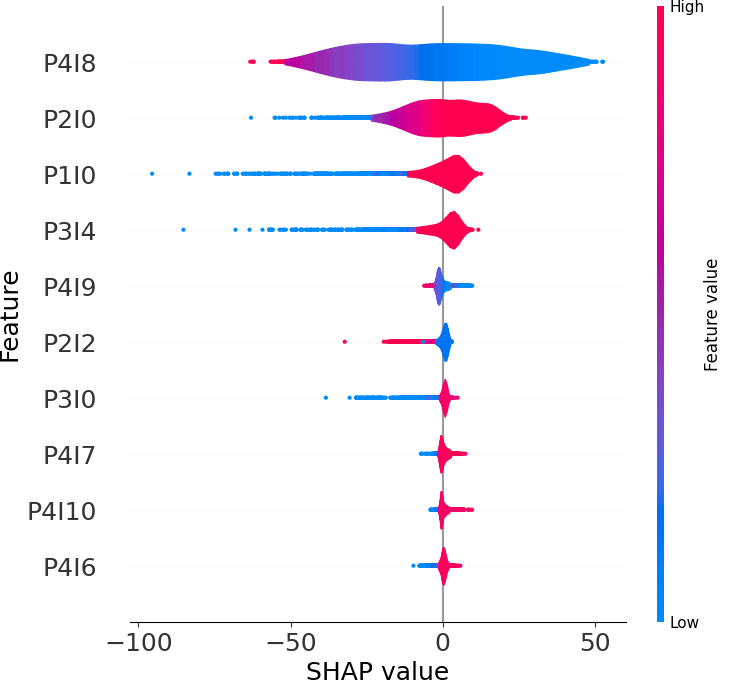}
    \caption{SHAP analysis of an Extra Trees regression, showcasing the first 10 SHAP values sorted by their impact on model output.}
    \label{fig:shap}
\end{figure}

Figure \ref{fig:shap} plots the SHAP analysis of an Extra Trees regression, showcasing the top 10 SF features sorted by their average SHAP value. 
A SHAP value indicates how much a given feature changes the output of the model compared to the baseline prediction; positive and negative SHAP values correspond to a positive and negative effect on the predicted property, respectively. The colors represent the values of the features for each data point; with red and blue values corresponding to high and low values of a particular feature, respectively. For example, if a dot is blue and located on the right side (positive SHAP value), the low value of that feature increases the value predicted by the model. 
The top five features from the SHAP analysis are also listed in Table \ref{tab:top5}, along with the sign of the correlation between the term and its effect on the predicted GB energy in that model.

\begin{figure*}[t]
    \centering
    \includegraphics[width=\textwidth]{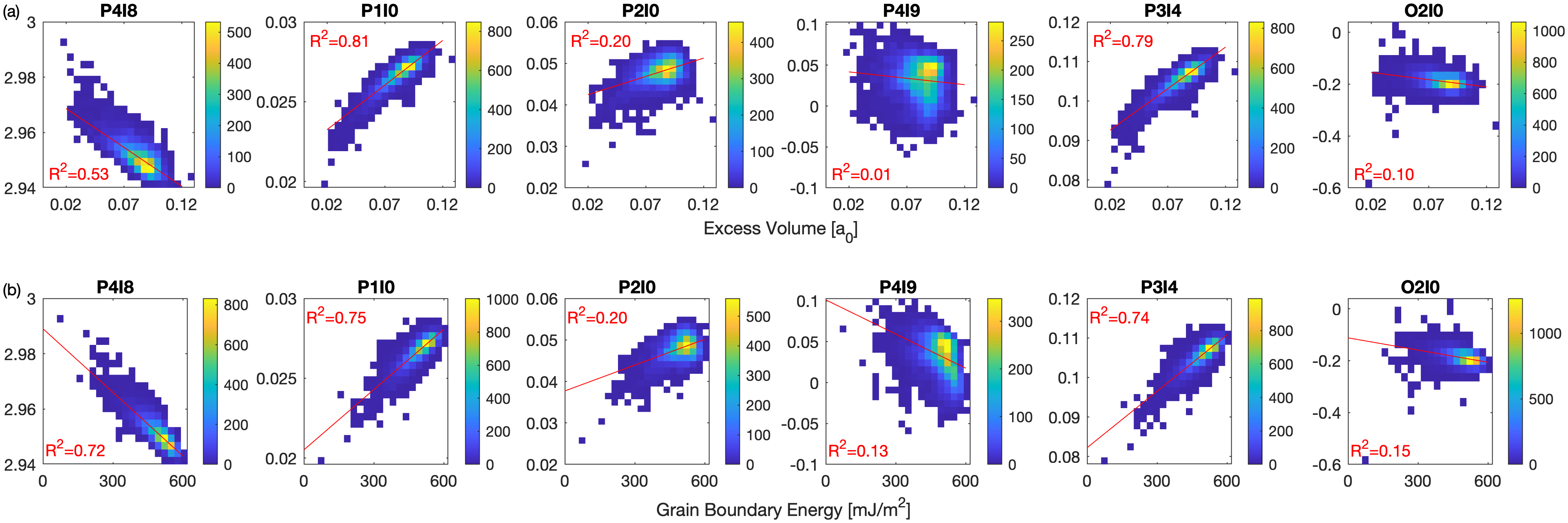}
    \caption{2D histograms of SF terms from Table \ref{tab:top5} with (a) excess volume, measured in lattice parameter units $a_0$, and (b) GB energy.}
    \label{fig:correlations}
\end{figure*}

There is significant alignment between the LASSO and SHAP analysis, as illustrated in Table \ref{tab:top5}. Four of the top five SF terms are the same in both analyses. Furthermore, the sign on the correlation of these four terms is the same. These four terms are P4I8, P1I0, P2I0, and P4I9, and are accompanied by the P3I4 and O2I0 terms that only appear in one model. 

To further confirm the expected correlation of these top terms from the LASSO and SHAP analysis, we plot the average SF values for each GB against both excess volume and energy in Figure \ref{fig:correlations}. Excess volume is included as it is known to have a positive correlation with GB energy \cite{homerExaminationComputedAluminum2022a}. It is noted that correlation plots for all SF terms are plotted in Supplemental Figures S8-S10. 
The top 5 features from these supplmental plots with the highest correlation with GB energy are listed in Table \ref{tab:top5} along with the sign of the correlation. It is worth noting that five of the six correlations plotted in Figure \ref{fig:correlations} have the same sign as that of the models listed in Table \ref{tab:top5}. The exception is the O2I0 term, which has an opposite sign in the model but only has a weak correlation with energy.

As noted earlier, SF comes with an added interpretability benefit since each SF descriptor classifies something unique about the deformation. Each SF term can be classified into one of five categories: density, deformation magnitude, deformation type, and internal and external orientation\footnote{Supplemental Figures S8-S10 
identify the categories for all 36 terms. Also, note that some of the categories from \cite{mishra2024learning} have been renamed: deformation magnitude was net deformation, deformation type was directionality, and the two orientation types were not listed as a categories.} \cite{mishra2024learning}.

The P4I8 term is one of three density metrics, and is an $r^4$ average for all the atoms in the neighborhood of a given atom. Figure \ref{fig:correlations} shows that there are reasonable correlations with excess volume and GB energy, which have$R^2$ values of $0.53$ and $0.72$, respectively.

The P1I0, P2I0, and P3I4 terms are all categorized as deformation magnitude.
P1I0 characterizes the gradient in the density. 
P2I0 characterizes the net deviatoric strain, akin to the von Mises strain invariant of the neighborhood. There should be substantial deformations of the fcc structures in the lattice neighboring the GB and shearing will be one of the primary means for this. The P3I4 term measures the extensional contribution of the strain gradient. The importance of this term can be attributed to the presence of strain gradients at GBs, especially in the case of GBs that can be represented as dislocation arrays. It is similar to the P1I0 term, where it is a gradient term, but it is weighted by $r^3$ rather than $r^1$. As illustrated in Figure \ref{fig:correlations}, the P1I0 and P3I4 have strong correlations with excess volume and energy, while P2I0's correlations are weak.

The P4I9 metric is an internal orientation metric that defines the orientation between the local lattice and the P4I6 measure of the shear. The latter is similar to the net shear metric P2I0, except that it is weighted by $r^4$ rather than $r^2$.  
Finally, the O2I0 is an external orientation metric that is also highlighted by the LASSO method; this terms defines the orientation of the lattice shearing (measured by P2I0) with respect to the normal of the GB. Thus, it has some similarity to P4I9. Given that aluminum is not isotropic (albeit with a relatively small Zener ratio), it is not surprising that the amount of shear necessary to accommodate the mismatch at a GB will be related to the direction of that shear i.e., the crystal orientation. 

Interestingly, the P4I9 and O2I0 terms both have little to no correlation with excess volume and energy on their own. But in conjunction with the other terms in the models, they are deemed more important for the prediction of GB energy than other terms that have strong correlation with energy. For example, the P2I2 term (a density metric based on $r^2$ weighting) has much stronger correlation with energy than the P4I9 term, but the models are not predicting based on any single term alone, but the combined effect of multiple terms. Thus, some of these terms may provide more of a secondary effect that can distinguish nuanced variations of the GB energy and such effects may not be apparent in 2-D cross-sections examining single-value correlations.

The consistency of the top features and their identifiable correlations with energy in all but one case illustrate that this combination of SF descriptor, average transform, and both linear and non-linear machine learning models is capturing useful trends. The average transform tells about general trends of the atoms in the GB but not about specific local atomic environments. The linear model provides detailed insight into the influence of each parameter. The non-linear model provides this insight through the SHAP analysis. Most importantly, the SF descriptor connects the features with physical, interpretable attributes of the GB structure-property relationship.

However, the analysis also uncovers the fact that different models will extract different features. While there are only minor differences in the top five features listed in Table \ref{tab:top5}, the difference becomes more dramatic as more terms are compared. In addition, other models will extract entirely different features; a Bayesian linear model applied to the data showed correlations with many of the SF metrics listed above, but sometime with the same effect and sometimes with an opposite effect on the GB energy. However, the Bayesian analysis, which is described in the supplemental material, required an additional step in describing the data (  subtracting the mean and dividing by the standard deviation). The model input is slightly different and leads to different correlations as a result. Care must be taken since the machine learning is only finding correlations in the data that has been through feature engineering, it is not finding the causal relationships for the predictions.

\section*{Discussion}
Prediction in variable-sized atom-clustered structures consistently require three steps: \textbf{describe}, \textbf{transform}, \textbf{machine learning}. Each of these steps, illustrated in Figure \ref{fig:flow}, play a role in the resulting model accuracy and interpretation. Such feature engineering is frequently employed to provide more accurate predictions or better interpretation of the results. In the fast growing environment of machine learning and artificial intelligence models, the diversity in steps taken by different groups makes it challenging to know which methods lead to improved accuracy and interpretability. We have examined this challenge in GBs, which, like all variable-sized atom clusters, require a transform to obtain consistent feature sizes. By attempting to standardize the various steps of the feature engineering process, we have aimed to understand how each step in the process affects the accuracy and interpretability of the resulting model predictions.

Descriptors play an essential role in taking atom structure information, most often represented by Cartesian coordinates, and mapping that to a feature vector that encodes the most critical information.
Our findings underscore the robustness of physics-inspired structural representations \cite{musilPhysicsInspiredStructuralRepresentations2021a} in capturing the intricate behaviors of GBs. Notably, descriptors such as the SOAP, ACE, and SF demonstrated superior predictive accuracy, underscoring their potential in advancing computational materials science. SF stands out in this group of accurate predictors because it has a low feature count and each feature has a physical meaning in terms of the strain in the neighborhood of each atom. Perhaps one of the interesting conclusions is that higher order deformations (i.e., strain gradients and higher) should be considered for accurate predictions of GB energy. This is likely the main reason behind higher predictive capability of SOAP and ACE compared to SF, as the current implementation of SF considers only up to 4\textsuperscript{th} order terms, whereas spherical harmonics up to 12\textsuperscript{th} order were considered for both SOAP and ACE. This is also corroborated by LASSO analysis for SOAP descriptors (Supplemental Figures S1 and S2), 
which shows persistence of $l=8,10$ terms.

In short, it appears that better accuracy can always be achieved with additional descriptor information. For example, there are numerous cases where concatenation of one or more descriptor improves the learning \cite{tamuraFastScalablePrediction2017,homerMachineLearningInformedRepresentations2019a,montesdeocazapiainCharacterizingTensileStrength2020}. But, longer feature vectors complicate interpretability. Additionally, some descriptors that can encode a lot of information cannot be processed in reverse to provide physical insight from model predictions. Therefore, those descriptors that encode physics that can be readily extracted from a model prediction are likely to provide the greatest insight into the resulting structure-property models.

The transforms applied to obtain consistent feature sizes from the variable-sized input data impact both the accuracy and interpretability of the resulting model. As hypothesized in this work, the average transform provides the best accuracy because it is the most similar to the procedure used to calculate GB energy from the atomic structure; the excess energy, relative to the bulk energy, for all the atoms is summed and divided by the area of the boundary. We further hypothesize that other properties of interest may benefit from descriptors that capture the important features of that property. For example, properties controlled by extreme values in a distribution may benefit from a transform that captures the structural aspects of those extreme values. However, while some transforms might provide more accurate predictions, they may make interpretation difficult. The average transform is one of these and it can not be processed in reverse to tell us how the distribution of values that were averaged might be critical to a certain structure-property relationship.

The machine learning models and algorithms also play a significant role in both accuracy and interpretability. As discussed above, simpler models are preferred to complex models and the linear models provide high accuracy in many cases. The linear models are easier to interpret because the contribution of any given feature can be easily discerned. However, non-linear models can make use of Explainable AI tools \cite{longoExplainableArtificialIntelligence2024}, such as SHAP analysis discussed above, to extract interpretable meaning from the resulting model. Models can sometimes be overly complex; we suggest opting for simpler linear models whenever feasible.

From an accuracy standpoint, we recognize that we have not considered all combinations of descriptors, transforms, and machine learning models. Neither have we done an exhaustive search of the hyperparameters for each of the descriptors, transforms and machine learning models beyond a simple gridsearch of hyperparameters for the machine learning models. We cannot guarantee that higher accuracy couldn't be achieved with adjustments to the models we examined. But, by providing a systematic approach, and a consistent dataset across all the models evaluated in this work, we provide a framework and benchmark against which new and improved models can be tested, like the MNIST dataset has served for benchmarking machine learning efforts in optical character recognition \cite{lecun-98,Deng:2012:MNIST,Bengio:2013:MLreview_MNIST}. Furthermore, by providing standard steps and language for the comparison of models along with a deliberate attempt to employ principles of the machine learning community, we hope the grain boundary community can identify the best methods to obtain structure-property relationships that will drive innovation.

Feature selection is a tool that can be used in conjunction with machine learning models to reduce the feature vector to those items that are the most critical for the accuracy in the model. The assumption from an interpretability standpoint is that these selected features are the most important for the model and can therefore be used to obtain insight into the structure-property models.

The final case study illustrated how the SF descriptor provides insight into the density and deformations that correlate with GB energy. Four of the top five features were shared between a linear and non-linear model.
This nuanced view, where some features are consistently highlighted across methods while others are unique to specific approaches, offers a richer understanding of the predictive landscape. It suggests that while some attributes of GBs are universally recognized by various predictive models, others may be more method-dependent, possibly due to underlying assumptions or mathematical formulations inherent to each technique. This insight not only enriches our understanding of feature selection dynamics but also guides further investigation into the specific roles these features play in material structure-property relationships.

The implications of these results offer a pathway towards more precise and efficient predictive models that can be instrumental in materials design and engineering. By enhancing our ability to predict GB properties, these findings could facilitate the development of materials with optimized mechanical properties, thereby having a profound impact on various industrial applications.

\section*{Methods}

In the application of machine learning grain boundaries (GBs), feature engineering \cite{hastieElementsStatisticalLearning2009} is crucial for enhancing model performance by tailoring input data to more accurately reflect the underlying problem and prepare for predictive modeling. Figure \ref{fig:flow} illustrates the three key steps taken for predicting GB properties: employing descriptors to encode or \textbf{describe} the structure (step 1), \textbf{transforms} to standardize the data size across different GBs (step 2), and  \textbf{machine learning} or prediction algorithms to predict a GB property from the input data (step 3). The methods and techniques employed in this work for the three key steps are described below.

\subsection*{Descriptors}

\begin{figure}
    \centering
    \includegraphics[width=\columnwidth]{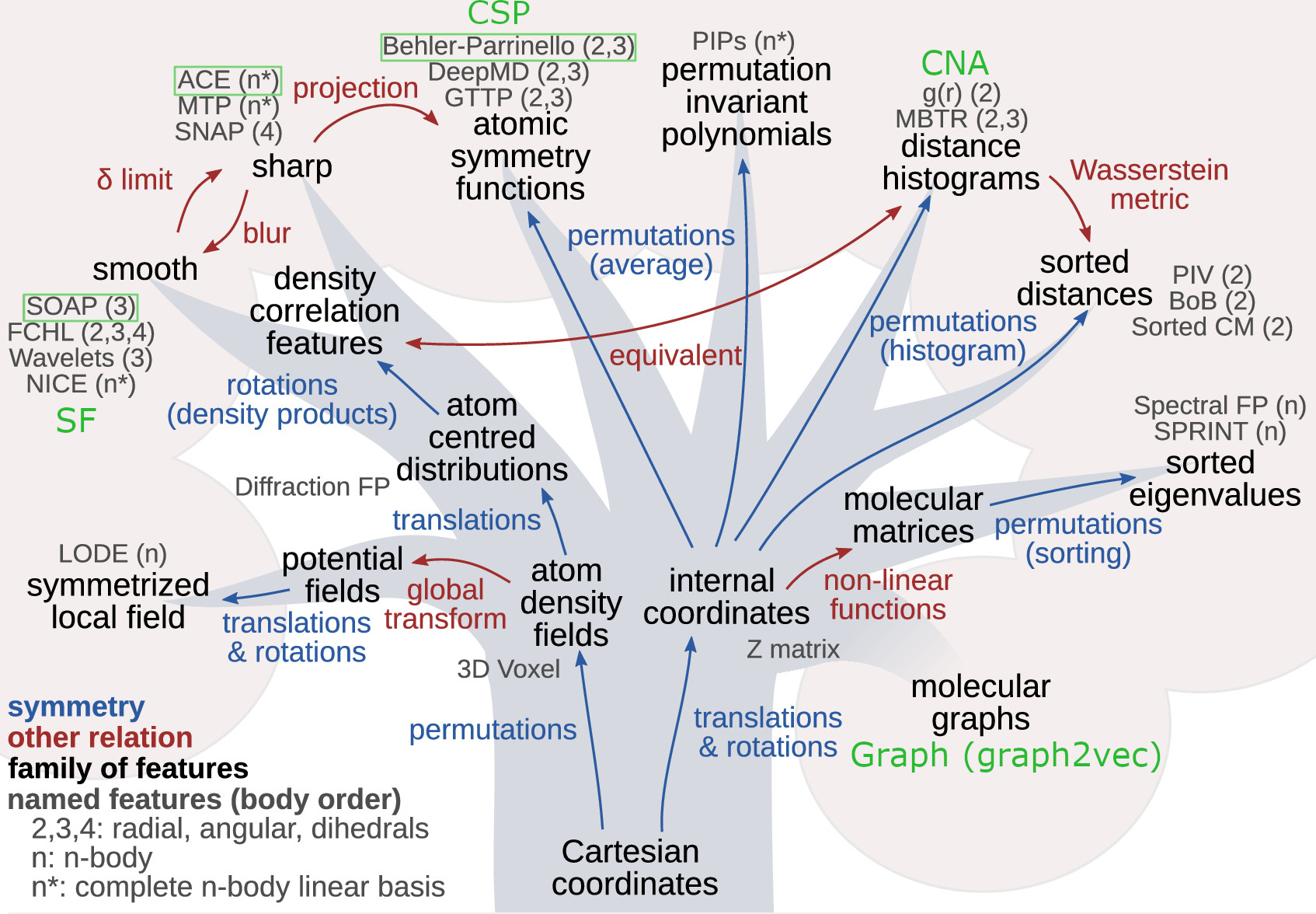}
    \caption{Philogenetic tree of structural representations adapted from Musil et al.\ \cite{musilPhysicsInspiredStructuralRepresentations2021a} under the CC-BY-4.0 license. All additions are marked in green to identify the descriptors used in the present work.}
    \label{fig:ourtree}
\end{figure}

The descriptors employed in this work are inspired by a phylogenetic tree of structural representations shown in Figure \ref{fig:ourtree}. The tree is reproduced and adapted from a paper titled ``Physics-Inspired Structural Representations for Molecules and Materials,'' by Musil et al.\  \cite{musilPhysicsInspiredStructuralRepresentations2021a}. They emphasized the requirements for developing descriptors that map the atomic positions in Cartesian coordinates to a new metric space commonly called the feature space, the representation \cite{musilPhysicsInspiredStructuralRepresentations2021a}, descriptor \cite{bartokRepresentingChemicalEnvironments2013a}, or fingerprint \cite{kearnesMolecularGraphConvolutions2016, sadeghiMetricsMeasuringDistances2013a, parsaeifardMaximumVolumeSimplex2020, parsaeifardManifoldsQuasiconstantSOAP2022, batraGeneralAtomicNeighborhood2019, piaggiEntropyBasedFingerprint2017, zhuFingerprintBasedMetric2016}. This mapping is required for predictive modeling because atom locations represented as Cartesian coordinates cannot uniquely characterise materials \cite{musilPhysicsInspiredStructuralRepresentations2021a}. Specifically, atom locations do not preserve symmetries and any translation or rotation of a material will change the Cartesian coordinates.
Thus, a descriptor should address requirements of completeness, symmetry, smoothness, and additivity. These requirements and the general process is discussed in additional detail in the supplemental materials.

In this work, we considered descriptors from all the branches. A more complete description of all the methods considered is provided in the supplemental materials. In the end, we selected the following methods listed in the tree by Musil et al.\  \cite{musilPhysicsInspiredStructuralRepresentations2021a}: smooth overlap of atomic positions (SOAP), atomic cluster expansion (ACE) \cite{drautzAtomicClusterExpansion2019}, and Behler-Parinello's atom centered symmetry functions (ACSF) \cite{behlerAtomcenteredSymmetryFunctions2011, behlerGeneralizedNeuralNetworkRepresentation2007}. These three methods are boxed in green in Figure \ref{fig:ourtree}. We also include several other methods. First, we include a recent Strain Functional (SF) Descriptor, which define the minimal set of invariants required to characterise deformations up to the 4\textsuperscript{th} order (i.e., second derivatives or curvature of the strain) \cite{koberStrainFunctionalsComplete2024}. This method is similar to SOAP in having smoothly declining neighborhood function, compared to the sharper neighborhood cutoffs of ACE, MTP and SNAP. Second, we include two methods commonly used to identify GB atoms: Centrosymmetry Parameter (CSP) and Common Neighbor Analysis (CNA). Finally, we include a periodic graph description of a boundary where every atom in the GB is defined by a node, which is connected to edges within a certain cutoff distance. Each of these additional methods is listed in Figure \ref{fig:ourtree} in green under the branch we identified as most relevant. Techniques considered but not selected from other branches are discussed in the supplemental materials, with reasons being that in some cases they were not appropriate for characterization of single species large GB structures and in other cases, code was not readily available for implementation.

\begin{table}[]
\centering
\caption{Table of output size of various descriptors}
\begin{tabular}{ll}
\hline
Descriptor         & M Value \\ \hline
ACE                & 121     \\
ACSF               & 37      \\
SOAP               & 1014    \\
SF & 36      \\
Graph              & n/a     \\
CSP                & 1       \\
CNA                & 1       \\ \hline
\end{tabular}
\label{tab:descriptor_input}
\end{table}

The SOAP, ACE, and ACSF implementations all required user defined parameters that impact how many terms is used to represent an atom environment. SOAP and ASCF are implemented using the `DScribe' python library \cite{himanenDScribeLibraryDescriptors2020}, a powerful tool designed for creating descriptors in atomistic systems. For SOAP, all default parameters were used excluding \texttt{rcut=3.74}, \texttt{lmax=12}, \texttt{nmax=12}, and \texttt{sigma=.575}. ACSF are unique in the sense that the user defines a list of interactions i.e. pairs, triplets for the various functions used by ACSF. Since these lists can not be described briefly here, they are included in the supplemental materials. The implementation of the ACE algorithm was achieved using the `ace.jl' package within the Julia programming environment \cite{drautzAtomicClusterExpansion2019, dussonAtomicClusterExpansion2021}. A basis was created using the following parameters: \texttt{N=3}, \texttt{maxdeg=12}, \texttt{r0=2.86}, \texttt{rin=0.1}, \texttt{rcut=3.25}, and \texttt{pin=2}. The SF descriptors are currently 4th order, but can be extended to higher orders. SF used the following parameters: \texttt{sigma=1.017837} and \texttt{cutoff=5.699887}. The graph description has the nodes of the periodic graph listed as their spatial coordinates. The edges are weighted by the distance to their neighbors that are within \texttt{rcut=3.74} \AA\ of the atom (node) of interest. CNA \cite{honeycuttMolecularDynamicsStudy1987} and CSP \cite{kelchnerDislocationNucleationDefect1998} both represent each atom's environment with a single value, which is calculated during the creation of the dataset using LAMMPS \cite{LAMMPS}. Table \ref{tab:descriptor_input} shows the feature lengths of each of our descriptors denoted by $M$.

\begin{table}[]
\caption{Table of input and output shapes of various transform methods and the parameter $P$ used to create the shapes}
\begin{tabular}{llll}
\hline
Transform       & Input          & Output       & $P$ Choice \\\hline
Average         & $N_i \times M$ & $1 \times M$ & n/a      \\
Largest Simplex & $N_i \times M$ & $P \times M$ & 10       \\
CUR             & $N_i \times M$ & $P \times M$ & 20       \\
Kmeans          & $N_i \times M$ & $P \times M$ & 100      \\
KDE             & $N_i \times 1$ & $P$          & 100      \\
graph2vec       & Graph          & $P \times M$ & 128      \\\hline
\end{tabular}
\label{tab:transform_input}
\end{table}

\subsection*{Transforms}
Since GBs and other variable-sized atom-clustered structures can have variable numbers of features, the feature size must be standardized through some sort of transform. This will transform the $N_i \times M$ feature representation from the descriptor to a $P \times M$ feature representation that is identical for all atomic structures.

In this work we examine 6 possible transforms :  average, CUR or skeleton matrices \cite{2020SciPy-NMeth}, \mytextsf{KMeans} clustering \cite{scikit-learn}, the largest simplex \cite{parsaeifardMaximumVolumeSimplex2020}, kernel density estimation (\mytextsf{GaussianKDE}) \cite{scikit-learn}, and \mytextsf{graph2vec} \cite{karateclub}.  The motivation behind this set of transforms, along with a more detailed explanation of each is provided in the supplemental materials.

The Average, largest simplex, CUR and \mytextsf{Kmeans} transforms are used with almost all descriptors. \mytextsf{GaussianKDE} is used only with CSP and CNA and graph2vec is used only with the graph descriptor. 
Table \ref{tab:transform_input} lists the theoretical input and output size of each transform along with the $P$ values used in this work. It is noted that no attempt was made to find the optimal $P$ value for each implementation. Explanations for why each $P$ value was chosen is given below along with the descriptions of the transforms.

\subsection*{Machine Learning}
Machine learning is oftentimes yet another mapping to a different feature space. To understand the impact of this step, the research employed a comprehensive and systematic approach to compare a diverse set of machine learning algorithms. This set includes three linear models: \mytextsf{LinearRegession}, \mytextsf{Lasso}, and \mytextsf{RidgeCV}; one support vector machine: \mytextsf{SVR}; two ensemble methods: Extra Trees and \mytextsf{AdaBoost}; one nearest neighbor method: \mytextsf{KNN}; and one neural network (deep learning) model: \mytextsf{MLPRegresssion}. All of these methods are available for implementation via the sklearn python library \cite{scikit-learn}. This set of algorithms were selected using the sklearn documentation where various supervised methods are grouped by methodology. 

Central to this methodology was the implementation of grid search for hyperparameter optimization across all algorithms. This was implemented in an effort to provide a uniform and equitable basis for comparison between models. 
We disclose that beyond the grid search, which required user input, little additional effort was taken for the optimization of the hyperparameter values. Typically convergence was costly from a time perspective for the non-averaging transforms due to the increased complexity, so the range of hyperparameters defined in the grid search was customised according to convergence time and not accuracy. While we recognize that this may limit the accuracy of these predictions, our data consisted of only a train and test set, so this also minimized the chances of overfitting.

A 5-fold cross-validation strategy was applied to all models.  
This approach assesses the models' performance on unseen data, providing a dependable estimate of their generalization capabilities. For performance assessment, both the Mean Absolute Error (MAE) and the Coefficient of Determination ($R^2$) were recorded. The MAE provided an understanding of the average absolute error made by the models, while the $R^2$ offered insights into the proportion of variance in the dependent variable that could be explained by the independent variables, acting as another method for measuring the accuracy of the model.

Comprehensive documentation of all aspects of the model training process, including hyperparameter values, cross-validation details, and performance metrics, was maintained for the reproducibility of the research. A code base with a sample dataset is included to ensure that the experiments can be replicated and validated by others in the scientific community.

\subsection*{Dataset}
To compare the diverse \textbf{descriptors}, \textbf{transforms}, and \textbf{machine learning} techniques, we employ a recently published atomistic dataset of aluminum GBs  \cite{homerExaminationComputedAluminum2022a, Homer:2022:AlGBdataset}. This dataset provides a comprehensive sampling of the five degrees of crystallographic character. The datasest includes 7304 pure Aluminum GB structures and their corresponding energies. The construction and process of obtaining the minimum energy structure for each GB is described in detail \cite{homerExaminationComputedAluminum2022a} and the structures are available for download \cite{Homer:2022:AlGBdataset}.
These GB structures were created with the emperical embedded atom model (EAM) potential created by Mishin et al.\ \cite{mishinInteratomicPotentialsMonoatomic1999}. Thus, the physics inherent to these structures is limited to the accuracy of this emperical potential and the methods used to construct the GBs.
For purposes of limiting data storage, each GB includes $\pm$ 15 \AA\ of atoms relative to the expected location of the GB. Because the size of some of the GBs in the original dataset were too large for some of the descriptor implementations, we use a subset of 7174 GBs, which are those GB structures that contain less than 35,000 atoms. It is expected that this dataset contains sufficient diversity both in crystallographic character and atomic structure to serve as a robust basis for the comparisons and interpretations provided in this work.

\section*{Acknowledgements}
This work was primarily supported by the U.S.\ National Science Foundation (NSF) under Award \#DMR-1817321. EMK and NM were supported by internal LANL LDRD funding (XX9A, XXG0).

\section*{Competing Interests}

All authors declare no financial or non-financial competing interests. 

\section*{Data Availability}

The datasets generated and/or analysed during the current study are available in the Mendeley Data repository, https://doi.org/10.17632/4ykjz4ngw \cite{Homer:2022:AlGBdataset,homerExaminationComputedAluminum2022a}.

\section*{Code Availability}

The underlying code for this study is available at 

\noindent \href{https://github.com/braxtonowens/gbcompare}{https://github.com/braxtonowens/gbcompare}. 

\section*{CRediT author statement}

C. Braxton Owens: Conceptualization, Methodology, Software, Formal analysis, Data Curation, Writing - Original Draft, Writing - Review \& Editing, Visualization;
Gus L. W. Hart, Eric R. Homer - Conceptualization, Methodology, Formal analysis, Writing - Original Draft, Writing - Review \& Editing, Visualization, Supervision, Funding acquisition
Tyce W. Olaveson: Conceptualization, Software, Formal analysis;
Nithin Mathew, Jacob P. Tavenner, Edward M. Kober, Garritt J. Tucker - all performed the following related to the Strain Functional analysis: Formal analysis, Writing - Review \& Editing;

 \bibliographystyle{elsarticle-num} 
 \bibliography{cas-refs}









\end{document}